\documentclass[12pt]{article}

\usepackage{amsmath}
\usepackage{amsfonts}
\usepackage{amssymb}
\usepackage{color}
\usepackage{etoolbox}
\usepackage{graphics} 
\usepackage[margin=3.3cm]{geometry}
\usepackage{url}
\usepackage{hyperref}
\usepackage{color}
\definecolor{refcolor}{RGB}{0,0,190}
\hypersetup{
    colorlinks,
    citecolor=refcolor,
    filecolor=refcolor,
    linkcolor=refcolor,
    urlcolor=refcolor
}
\usepackage[round]{natbib}

\makeatletter 
\renewcommand\@biblabel[1]{}
\makeatother

\def\({\left(}
\def\){\right)}
\newcommand{\R}{\mathbb{R}}

\newcommand{\de}{\textnormal{d}}
\newcommand{\flrw}{Friedmann-Lema\^itre-Robertson-Walker}
\newcommand{\FLRW}{FLRW}
\newcommand{\schw}{Schwarzschild}
\newcommand{\rn}{Reissner-Nordstr\"om}

\newcommand{\ds}{\displaystyle}
\newcommand{\dsfrac}[2]{\ds{\frac{#1}{#2}}}

\begin{document}

\title{Spacetime Singularities and Invariance}

\author{
  O. Cristi Stoica\thanks{Horia Hulubei National Institute for Physics and Nuclear Engineering, Department of Theoretical Physics, Bucharest. Corresponding author: cristi.stoica@theory.nipne.ro}
  \and 
  Iulian D. Toader\thanks{The Research Institute, University of Bucharest; Descartes Centre for the History and Philosophy of
the Sciences and the Humanities, Utrecht University. Contact: itoad71@gmail.com or i.d.toader@uu.nl}
}

\date{}

\maketitle

\section{Introduction}

Spacetime singularities are an important topic in general relativity and in cosmology, but understanding their philosophical significance is rather a side-issue in contemporary debates in philosophy of physics. The prevailing attitude still seems to be that singularities constitute a breakdown of physical laws.\footnote{See, e.g., this remark in a recent collection on the philosophy of general relativity: ``If you think you have a singularity, then you can't use it in a physical model. You don't know how to include such an object in a physical system, either as the outcome of gravitational collapse or as an object that might affect other objects with its gravitational field. [...] you are paralyzed by incomprehension.'' \citep{schutz2012thoughtsGR}} A notable exception to this rather unfortunate state of affairs is John Earman's book, \textit{Bangs, Crunches, Whimpers, and Shrieks} \citep{earman1995bangs}, published 20 years ago, focused on issues related to the definition, proper characterization, and existence of singularities, as well as on several problems and hypotheses that they are thought to have given rise to. Proposing a tolerant attitude towards singularities, Earman discusses cosmic censorship, supertasks, and the horizon problem, among other issues. What we discuss in the present paper is a novel approach to singularities, based on a recent extension of general relativity that shows why singularities do not actually constitute a breakdown of physical laws: it is not only the case that physical laws are valid, but they also remain invariant at singularities \citep{Sto13a}. We are interested here in describing this kind of invariance, as well as in drawing its consequences for our understanding of equivalence in general relativity. In particular, adopting a distinction recently introduced by Dennis Dieks \citep{dieks2006GeneralCovariance}, we point out that the difference between the metrics at singularities and those outside of singularities is factual, rather than nomological, and that this justifies the extension of the principle of equivalence to singularities.

Singularities, let us recall, have been discovered in the simplest solution representing the spacetime outside a body --- the solution given by Schwarzschild, soon after Einstein proposed general relativity, one hundred years ago. As typically characterized, singularities are regions where the metric tensor is no longer regular, so that the mathematical objects in the Einstein equation and other field equations become infinite or undefined. Most physicists, in particular Einstein, initially rejected the possibility of singular solutions. The hope was that they are due to idealizations like the perfect spherical symmetry, and would not occur in the real world, but the singularity theorems by Penrose \citep{Pen65,Pen69} and Hawking \citep{Haw66i,Haw66ii,Haw67iii,HP70} show that they are in fact unavoidable. Singularities are indeed predicted by the theory to occur both inside black holes, and at the Big Bang. This confirmed the ideas of those who thought that general relativity contained the seeds of its own destruction. Penrose then proposed the cosmic censorship hypothesis, which states that although singularities exist, they are isolated beyond the event horizon and so don't affect the physics outside the black hole \citep{Pen79,Pen98}. Hawking, however, showed that if quantum effects are considered, black holes can evaporate, and so the problem persists and it is even aggravated \citep{Haw75,Haw76}. This is the well-known information loss paradox.

The novel approach to singularities, adopted here and briefly described further below, gives alternative formulations of the geometric objects and the field equations, which don't break down at singularities and also remain invariant. A new branch of geometry emerges in this way, namely an invariant and more general extension of singular semi-Riemannian geometry \citep{Sto11a}, as well as a new physical theory: \textit{singular general relativity} \citep{Sto13a}. This provides an alternative formulation of differential geometry and general relativity, one which is equivalent to the standard one, but which can be extended at singularities too. This formulation turns out to solve many of the problems related to singularities. But its philosophical significance has yet to be fully articulated and evaluated.

Famously, Einstein and Rosen had remarked that one could multiply the standard equations by a suitable quantity, which vanishes at the singularity and removes the infinities, so the singularities may be not harmful after all \citep{ER35}. The novel approach adopted here can be seen as a mathematical follow-up on this remark, in the sense that it provides an invariant account leading to this multiplication, and justifies it mathematically and physically. This approach not only commends an attitude of tolerance towards singularities \citep{earman1996tolerance}, but is animated by the belief that, when correctly understood, they are a source of fruitful developments in general relativity and quantum gravity \citep{Sto12d}.

As far as we can see, the novel approach to spacetime singularities has important consequences for several topics of interest in philosophy of physics. Since at singularities, the distance between distinct points, as well as the duration between distinct events, can vanish, a further revision of our notion of spacetime, already revised by relativity and quantum mechanics, seems to be required. Similarly, issues concerning causality arise as well. Since distinct events are no longer separated in space or time, how does that change the way we should think about causality? Can the evolution equations be extended beyond the singularity? Are singularities really the ``end of the line,'' as is often believed? Standard treatments of singularities don't seem to help here, but the new approach allows the extension of spacetime and the fields beyond singularities. There are also related issues, having to do with the occurrence of Cauchy horizons, which seem to accompany charge and rotating black holes. Cauchy horizons causally disconnect some regions of spacetime, in the sense that the evolution equations defined on spacelike surfaces in the past of these regions don't reach them. The new approach to singularities allows black hole solutions to be compatible with global hyperbolicity, and hence with the absence of Cauchy horizons \citep{Sto12e}.

Furthermore, formulating general relativity and the geometry of spacetime in a way which is defined at singularities seems to suggest that the variables used in this formulation are more fundamental than the standard ones, which fail at singularities. This raises a metaphysical question about fundamentality: what geometric and physical fields are really the fundamental ones, and why? In particular, some charts are singular and are not good for representing the black hole solutions around the event horizons, but others are not singular, and are proper for this task. So the choice of the differential structure of the spacetime manifold can be done in many ways, but this choice should receive a good physical, mathematical, and philosophical justification.

In this paper, we want to focus our discussion on the problem of invariance of the physical laws. Since at singularities the metric is no longer regular, but degenerate, Lorentz invariance (in the tangent space at each spacetime point) is violated and needs to be replaced by a different type of invariance. But what type, more precisely?

\section{Singularities in General Relativity}

According to special relativity, physical laws are invariant with respect to the Poincar\'e group \citep{einstein1905elektrodynamik}. This is the group of isometries of the four-dimensional spacetime (the Minkowski spacetime), and includes translations and linear transformations that leave invariant the Lorentz metric
\begin{equation}
\label{eq:lorentz_metric_minkowski}
\Delta s^2 = - c^2\Delta t^2 + \Delta x^2 + \Delta y^2 + \Delta z^2,
\end{equation}
where $t$ is the time coordinate, $x,y,z$ are Cartesian space coordinates, and $c$ is the speed of light. In the absence of gravitational forces, the equations expressing physical laws preserve their form when a Poincar\'e transformation is applied. In special relativity, the set of positions occupied at different moments of time by an inertially-moving particle represents a straight line. This line can be parametrized so as to give its \emph{proper time}. To every instant of time corresponds a point on the time axis of the observer, and a three-dimensional space orthogonal (with respect to the inner product \eqref{eq:lorentz_metric_minkowski}) to the time axis at that point gives the \emph{proper space} of the observer. An orthogonal \emph{inertial reference frame} of the observer can be defined by choosing an origin on the time axis, a future-pointing vector on the time axis, and an orthogonal frame in the three-dimensional space orthogonal to the time axis at the origin. All such reference frames are related to one another by a Poincar\'e transformation.

Of course, it is not necessary for a reference frame to consist of orthogonal vectors. Any affine transformation of the Minkowski spacetime can transform any inertial frame into another inertial frame, but only the Poincar\'e transformations preserve the orthogonal character of inertial frames. In order to deal with inertial frames that are not orthogonal, one has to change the form of the metric, although the metric itself is an invariant mathematical object. The general form of the metric, valid in any inertial frame, is 
\begin{equation}
\label{eq:lorentz_metric_minkowski_nonorthogonal}
\Delta s^2 = \sum_{a,b=0}^3g_{ab}\Delta x^a \Delta x^b = g_{ab}\Delta x^a \Delta x^b,
\end{equation}
where $(x^0,x^1,x^2,x^3) := (t,x,y,z)$. By Einstein's summation convention, one can drop the $\sum_{a,b=0}^3$ symbol and sum over all indices which appear twice, in both lower and upper positions. In the case of non-inertial observers, one has to use curvilinear coordinates, such as the spherical coordinates, where the form of the metric tensor varies from point to point and the quantities $\Delta x^a$ become infinitesimal. This means that while for orthogonal inertial frames we could identify the Minkowski spacetime with a four-dimensional vector space on which the metric is defined, for curvilinear coordinates one can no longer make this identification. Instead, one introduces at each point of spacetime a four-dimensional vector space, the so-called \emph{tangent space} to the Minkowski spacetime at that point. The metric is then actually defined on the tangent space at that point. So, the infinitesimal length becomes
\begin{equation}
\label{eq:metric}
\de s^2 = g_{ab}\de x^a \de x^b,
\end{equation}
where the coefficients $g_{ab}$ are taken to depend on position.

Observing that curvilinear coordinates introduce additional inertial forces, Einstein realized that, if spacetime is curved, gravity can be included as such an inertial force, thereby obtaining the theory of general relativity. But, of course, on a curved spacetime the notion of Poincar\'e transformations no longer applies. Spacetime can no longer be identified with a vector space, and the Lorentz metric is no longer the same in all spacetime. Due to the curvature, one can use only local curvilinear coordinates, defined on open neighborhoods, and the metric varies from point to point, being defined independently on the tangent space at each point of the spacetime manifold. The coordinate transformations are local diffeomorphisms. On the Minkowski spacetime, diffeomorphisms include the Poincar\'e transformations. The field equations, that is, the Einstein equation and the equations describing the behavior of matter fields, are all formulated in tensor formalism, and are invariant under the group of diffeomorphisms. The Einstein equation is
\begin{equation}
\label{eq:einstein}
	G_{ab} + \Lambda g_{ab} = \kappa T_{ab},
\end{equation}
where the Einstein tensor $G_{ab}=R_{ab}-\dsfrac 1 2 Rg_{ab}$ is equated with the stress-energy tensor $T_{ab}$ of matter, and the constant $\Lambda$ is the cosmological constant, responsible for the accelerated expansion of the universe. This formulation does not violate Lorentz invariance, because diffeomorphisms also change the components of the metric tensor, so lengths remain invariant under diffeomorphisms. Therefore, in general relativity, Lorentz invariance is preserved in the tangent space at each point in spacetime.

However, as is well known, soon after Einstein gave this formulation of the equation, {\schw} found a solution describing the spacetime for a spherically symmetric gravitational field. The {\schw} metric is, in {\schw} coordinates,
\begin{equation}
\label{eq_schw_schw}
\de s^2 = -\(1-\dsfrac{2m}{r}\)\de t^2 + \(1-\dsfrac{2m}{r}\)^{-1}\de r^2 + r^2\de\sigma^2,
\end{equation}
where
\begin{equation}
\label{eq_sphere}
\de\sigma^2 = \de\theta^2 + \sin^2\theta \de \phi^2.
\end{equation}
It turns out that this metric has two singularities, one corresponding to $r=2m$ (the so-called \emph{event horizon}), and another to $r=0$. As showed by Eddington \citep{eddington1924comparison} and Finkelstein \citep{finkelstein1958past}, suitable coordinate transformations can remove the event horizon singularity, proving that this is an artefact of the {\schw} coordinates. But the same does not work for the $r=0$ singularity, because no coordinate transformations can remove the infinite value of the Kretschmann scalar $R^{abcd}R_{abcd}$ at $r=0$. Another exact solution was proposed by Friedmann, representing an expanding universe \citep{FRI22de,FRI24,FRI99en}. Its modern form is the {\flrw} (\FLRW) metric,
\begin{equation}
	\de s^2 = -\de t^2 + a^2(t)\de\Sigma^2,
\end{equation}
where
\begin{equation}
\label{eq_flrw_sigma_metric}
\de\Sigma^2 = \dsfrac{\de r^2}{1-k r^2} + r^2\(\de\theta^2 + \sin^2\theta\de\phi^2\).
\end{equation}
The spacetime manifold is here the product between a one-dimensional manifold representing the time, and a three-dimensional \emph{symmetric space} $\Sigma$, which can be the three-sphere $S^3$, the Euclidean space $\R^3$, or the hyperbolic space $H^3$. Also $k=1$ for $S^3$, $k=0$ for $\R^3$, and $k=-1$ for $H^3$. This solution has one singularity, at the \emph{big-bang}, where $a(t)=0$.

\section{Singular General Relativity}

The fact that these two solutions have singularities is not an accident due to the very high symmetry, as one initially hoped. The theorems offered by Penrose \citep{Pen65,Pen69} and Hawking \citep{Haw66i,Haw66ii,Haw67iii,HP70} show that, in general relativity, singularities have to occur in very general situations, such as encountered in our universe. The $r=0$ singularity of the {\schw} black hole is a spacelike singularity: time seems to end there. Any matter falling into the black hole reaches the singularity and vanishes, together with the information which it contains. The problem only gets bigger when quantum theory is taken into account. Because of particle creation in curved spacetime \citep{Haw75}, black holes evaporate, and at the end, part of the information describing the state of the universe appears to be lost \citep{Haw76}. But what happens to Lorentz invariance? 

Before we answer this question, let's distinguish between \emph{benign} singularities and \emph{malign} singularities, corresponding to the two main ways in which the metric tensor $g_{ab}$ can become singular: by having all its components smooth, but vanishing determinants, so that the metric becomes degenerate, and by having some of its components become infinite, respectively. In the case of a degenerate metric, the reciprocal metric $g^{ab}$ is singular. This makes it impossible to use two important tensor operations: the contraction between covariant components, which is usually done by contracting with $g^{ab}$, and the covariant derivative, which is typically given in terms of the Christoffel symbol of the second kind,
\begin{equation}
\label{eq_christoffel_second_kind}
		\Gamma^{c}{}_{ab} = \ds{\frac 1 2} {g^{cs}}(
		\partial_a g_{bs} + \partial_b g_{sa} - \partial_s g_{ab}).
\end{equation}
Without the covariant derivative, in particular, one can neither write the field equations, nor can one define the Riemann curvature tensor, $R^a{}_{bcd}$, in the usual way. In the case of a degenerate metric that has a constant signature \citep{Barb39,Moi40,Str41,Str42a,Str42b,Str45,Vra42}, one can define an operation similar to the covariant derivative \citep{Kup87a,Kup87b,Kup87c,Kup96}. However, the definition relies on choosing a subspace on the tangent space at each point, in a non-invariant way. Moreover, it could not be used for spacetime singularities, due to the constant signature of the metric.

More recently, however, one of the authors developed a novel approach that works, in an invariant way, with degenerate metrics that have a variable signature \citep{Sto11a}. The problem of contraction between covariant components has been resolved for a special type of tensors, which belong to tensor products of the tangent space and a subspace of the cotangent space of covectors of the form $g_{ab}v^b$. It turns out that such tensors contain all that is needed to describe singularities in terms of finite quantities, and even to construct the Riemann curvature. This approach defines a different kind of covariant derivative, which remains finite, in terms of the Christoffel symbol of the first kind,
\begin{equation}
\label{eq_christoffel_first_kind}
		\Gamma_{abc} = \ds{\frac 1 2} (\partial_a g_{bc} + \partial_b g_{ca} - \partial_c g_{ab}).
\end{equation}
This allowed the definition of the Riemann curvature tensor $R_{abcd}$. As a consequence, Einstein's equation \eqref{eq:einstein} could be rewritten, in a different form, which is however equivalent to \eqref{eq:einstein} outside the singularities, but remains finite and smooth at singularities \citep{Sto11a,Sto12b}. The new formalism can handle the {\FLRW} singularities as well, which turned out to be benign, resulting in a geometric and physical description in terms of finite quantities \citep{Sto11h,Sto12a}. Similarly, for the black hole singularities, coordinate transformations were found that transform the malign singularity at $r=0$ into a benign one \citep{Sto11e,Sto11f,Sto11g}. In the case of the {\schw} singularity, it has been shown that the coordinate transformation
\begin{equation}
\label{eq_coordinate_semireg}
\begin{array}{l}
\bigg\{
\begin{array}{ll}
r &= \tau^2 \\
t &= \xi\tau^4. \\
\end{array}
\\
\end{array}
\end{equation}
results in the following form of the metric
\begin{equation}
\label{eq_schw_analytic_tau_xi}
\de s^2 = -\dsfrac{4\tau^4}{2m-\tau^2}\de \tau^2 + (2m-\tau^2)\tau^{4}\(4\xi\de\tau + \tau\de\xi\)^2 + \tau^4\de\sigma^2,
\end{equation}
which renders the $r=0$ singularity benign. The {\schw} singularity is, of course, still a singularity, because the metric is degenerate, but the new formalism can be applied to obtain a geometric description of the singularity in terms of finite quantities. Also, the solution extends analytically beyond the singularity, so that in the case of Hawking evaporation, the spacetime is recovered. The upshot of the new approach is that, if they are benign, spacetime singularities do not constitute a problem for our physical theorizing, and in particular they should not be considered as a breakdown of physical laws. 

Naturally, at those points in the spacetime manifold where the metric is degenerate, Lorentz invariance is violated \textit{because} the metric is degenerate. On account of this violation, the Lorentz group has to be replaced by a group of linear transformations of the tangent space that preserve the degenerate metric at those very points. But how does such a group look like?

A degenerate metric $g$ on a vector space $V$, in particular on the tangent space at a spacetime point, defines an inner product on $V$, by $\langle u,v\rangle = g_{ab}u^a v^b$. The vectors $v\in V$ satisfying $\langle u,v\rangle=0$ for any vector $u\in V$ are called degenerate, and they form a vector subspace of $V$, called the \emph{radical} of $g$. The vector space $V$ can be split as a direct sum between the radical and a complementary vector space of dimension equal to the rank of the metric, on which the restriction of $g$ is non-degenerate. The vectors from the dual space $V^\ast$ act on $V$ as linear forms. Those co-vectors which vanish on the radical of $g$ form a vector space of dimension equal to the rank of the metric, called the \emph{annihilator} of the radical. A general linear transformation on $V$ also acts on the dual $V^\ast$. If the linear transformation preserves the metric, it has to map the radical onto itself. On the dual space, it maps the annihilator onto itself while preserving the non-degenerate metric induced by $g$ on the annihilator. These general linear transformations form a group which preserves the metric. 

Such groups were studied by the Romanian mathematician Dan Barbilian \citep{Barb39}. In particular, if the metric is non-degenerate, the group is the orthogonal group of the metric. If the metric is degenerate in all directions, that is, if it vanishes, then the group preserving it is the general linear group of $V$. The group that is needed, in our case, to replace the Lorentz group at singularities is the intermediate case, i.e., a group that preserves the degenerate metric only at some points in spacetime. The replacement is not required outside the singularities, where the metric is non-degenerate. Thus, outside the singularities there is no violation of the Lorentz invariance (in the tangent space at every point). The replacement, moreover, does not have any physical consequences outside the singularities (at least if quantum effects are ignored).

\section{Discussion}

Now, in summary, the standard formulation of general relativity presents the Einstein equation as valid and generally covariant outside the singularities, but as breaking down at singularities. However, for the reasons expounded above, it turns out that this standard formulation can be circumvented. The extension of general relativity to singularities introduces an equivalent, generally covariant reformulation of the Einstein equation, which does not break down at (benign) singularities. Its fundamental equation, just like the Einstein equation, is invariant under the symmetry group of local diffeomorphisms. Unlike the Einstein equation, the fundamental equation of singular general relativity is also Barbilian invariant, i.e., invariant under a symmetry group that preserves only the degenerate metric at (benign) singularities, in the tangent space at every point.

Furthermore, singular general relativity also extends the principle of equivalence to singularities. In a recent paper, Dennis Dieks introduced an useful distinction between factual and nomological differences between reference frames, i.e., ``differences that should be seen as \textit{fact-like} rather than \textit{law-like}'' \citep{dieks2006GeneralCovariance}. He argued that, whereas in classical mechanics and in special relativity, the differences between inertial and accelerated frames should be understood as nomological differences, in general relativity, the differences between all frames should be understood as purely factual differences, since the spacetime manifold itself is a dynamical structure rather than a fixed background. This allows for an understanding of the equivalence between reference frames in the sense that there are no nomological differences between them, and justifies the claim, which had been repeatedly challenged, that Einstein was right in thinking that general relativity extends the principle of equivalence to accelerated motion. In other words, understanding equivalence in the way suggested by Dieks justifies the claim that Einstein was right in thinking that the equivalence of all arbitrary frames is a consequence of the general covariance of his equation. This only assumes that general covariance is regarded not as a merely formal requirement, in the sense that the Einstein equation should possess the same syntactic form under arbitrary coordinate transformations \citep{norton1995,earman2006}, but rather as a substantive requirement that the same laws hold in all reference frames, that is, that they contain no quantities locating the frame with respect to a fixed spacetime background. On this interpretation of general covariance, all arbitrary frames can be equivalent despite being factually distinct.

We agree with this interpretation of general covariance, since we think it true that the metric itself is a dynamical object, rather than a fixed background, as it is in classical mechanics and in special relativity. But, furthermore, we believe that distinction introduced by Dieks can be used in supporting the view that the principle of equivalence, understood in nomological terms, should be extended not only to accelerated motion, but to spacetime singularities as well. For if the same laws that hold in all reference frames are also valid at singularities, and if differences of metrics are factual rather than nomological, then the principle of equivalence can be extended to singularites. As described above, it is indeed the case that the same laws that hold outside of singularities are also valid at singularities, and it seems natural to take the degeneracy of the singular metric as a matter of fact, rather than as a matter of law. So the principle of equivalence can be extended farther than Einstein thought it would. 

But some may want to deny that the degeneracy of the singular metric is a matter of fact. However, besides the dynamical character of the metric, which we take to strongly suggest that its degenerate character is factual rather than nomological, here is another reason that could justify this idea. It is known that the metric tensor of a distinguishing spacetime can be obtained from the causal structure (the collection of lightcones) up to a scaling factor \citep{Zeeman1964CausalityLorentz,Zeeman1967TopologyMinkowski,Malament1977TopologySpaceTime}. The scaling factor can be recovered by knowing a measure, which gives the volume form \citep{Finkelstein1969SpaceTimeCode}. Therefore, at the topological level, ignoring the differential structure, the metric can be expressed equivalently by the causal structure and the measure, which are topologically invariant. It has been shown that the topology of the lightcones is the same at singularities as it is outside them, at least for the {\flrw} big-bang singularity and for the {\schw} and {\rn} black hole singularities \citep{Sto15b}. The measure is also not manifestly special at these singularities. But if these two structures -- the causal structure and the measure, which again depend only on the topology -- are not manifestly special at singularities, then how does the degeneracy of the metric arises? The reason is that the differential structure ``arranges'' the spacetime events on the manifold in a certain way, and forces the metric to be degenerate in some cases. With respect to the causal structure, the lightcones become flattened in some directions of spacetime, although topologically they are equivalent. At the topological level the local diffeomorphisms are replaced by local homeomorphisms. So, while general covariance still preserves the degenerate or regular character of the metric, an extension of general covariance, where local diffeomorphisms are replaced by local homeomorphisms, obliterates any differences between the degenerate and regular metrics, which further supports the idea that the character of the metric is factual rather than nomological. However, extending the field equations, including the Einstein equation, so that they are invariant to homeomorphisms and not merely to diffeomorphisms, is of course still an open problem. The difficulty is due to the fact that all field equations are partial differential equations, which make sense for our current mathematical understanding only in the presence of a differential structure. Nevertheless, it is safe to conclude that even limited to local diffeomorphisms, the laws are the same outside the singularities and at singularities, and the differences between frames are only factual, which reinforces our point above that the principle of equivalence, interpreted in nomological terms, can be extended to singularities.\footnote{The Research Institute at the University of Bucharest, where the paper has been finalized, is gratefully acknowledged by the second author for financial and institutional support.}

\bibliographystyle{chicago2}

\begin{thebibliography}{}

\bibitem[\protect\citeauthoryear{Barbilian}{Barbilian}{1939}]{Barb39}
Barbilian, D. 1939.
\newblock {Galileische Gruppen und quadratische Algebren}.
\newblock {\em Bull. Math. Soc. Roumaine Sci.\/}, XLI.

\bibitem[\protect\citeauthoryear{Dieks}{Dieks}{2006}]{dieks2006GeneralCovariance}
Dieks, D. 2006.
\newblock Another look at general covariance and the equivalence of reference
  frames.
\newblock {\em Stud. Hist. Philos. Sci. B: Stud. Hist. Philos. M. P.\/}~{\em
  37\/}(1), 174--191.

\bibitem[\protect\citeauthoryear{Earman}{Earman}{1995}]{earman1995bangs}
Earman, J. 1995.
\newblock {\em Bangs, Crunches, Whimpers, and Shrieks-Singularities and
  Acausalities in Relativistic Spacetimes}.
\newblock {Oxford University Press, USA}.

\bibitem[\protect\citeauthoryear{Earman}{Earman}{1996}]{earman1996tolerance}
Earman, J. 1996.
\newblock Tolerance for spacetime singularities.
\newblock {\em Found. Phys.\/}~{\em 26\/}(5), 623--640.

\bibitem[\protect\citeauthoryear{Earman}{Earman}{2006}]{earman2006}
Earman, J. 2006.
\newblock Two challenges to the requirement of substantive general covariance.
\newblock {\em Synthese\/}~{\em 148\/}(2), 443--468.

\bibitem[\protect\citeauthoryear{Eddington}{Eddington}{1924}]{eddington1924comparison}
Eddington, A.~S. 1924.
\newblock {A Comparison of Whitehead's and Einstein's Formulae}.
\newblock {\em Nature\/}~{\em 113}, 192.

\bibitem[\protect\citeauthoryear{Einstein}{Einstein}{1905}]{einstein1905elektrodynamik}
Einstein, A. 1905.
\newblock Zur elektrodynamik bewegter k{\"o}rper.
\newblock {\em Annalen der Physik\/}~{\em 322\/}(10), 891--921.

\bibitem[\protect\citeauthoryear{Einstein and Rosen}{Einstein and
  Rosen}{1935}]{ER35}
Einstein, A. and N.~Rosen 1935.
\newblock {The Particle Problem in the General Theory of Relativity}.
\newblock {\em Phys. Rev.\/}~{\em 48\/}(1), 73.

\bibitem[\protect\citeauthoryear{Finkelstein}{Finkelstein}{1958}]{finkelstein1958past}
Finkelstein, D. 1958.
\newblock Past-future asymmetry of the gravitational field of a point particle.
\newblock {\em Phys. Rev.\/}~{\em 110\/}(4), 965.

\bibitem[\protect\citeauthoryear{Finkelstein}{Finkelstein}{1969}]{Finkelstein1969SpaceTimeCode}
Finkelstein, D. 1969.
\newblock Space-time code.
\newblock {\em Physical Review\/}~{\em 184\/}(5), 1261.

\bibitem[\protect\citeauthoryear{Friedman}{Friedman}{1922}]{FRI22de}
Friedman, A. 1922.
\newblock {{\"U}ber die Kr{\"u}mmung des Raumes}.
\newblock {\em Zeitschrift f{\"u}r Physik A Hadrons and Nuclei\/}~{\em
  10\/}(1), 377--386.

\bibitem[\protect\citeauthoryear{Friedman}{Friedman}{1924}]{FRI24}
Friedman, A. 1924.
\newblock {{\"U}ber die M{\"o}glichkeit einer Welt mit konstanter negativer
  Kr{\"u}mmung des Raumes}.
\newblock {\em Zeitschrift f{\"u}r Physik A Hadrons and Nuclei\/}~{\em
  21\/}(1), 326--332.

\bibitem[\protect\citeauthoryear{Friedman}{Friedman}{1999}]{FRI99en}
Friedman, A. 1999.
\newblock {On the Curvature of Space}.
\newblock {\em Gen. Relat. Grav.\/}~{\em 31\/}(12), 1991--2000.

\bibitem[\protect\citeauthoryear{Hawking}{Hawking}{1966a}]{Haw66i}
Hawking, S.~W. 1966a.
\newblock The occurrence of singularities in cosmology.
\newblock {\em P. Roy. Soc. A-Math. Phy.\/}~{\em 294\/}(1439), 511--521.

\bibitem[\protect\citeauthoryear{Hawking}{Hawking}{1966b}]{Haw66ii}
Hawking, S.~W. 1966b.
\newblock The occurrence of singularities in cosmology. {II}.
\newblock {\em P. Roy. Soc. A-Math. Phy.\/}~{\em 295\/}(1443), 490--493.

\bibitem[\protect\citeauthoryear{Hawking}{Hawking}{1967}]{Haw67iii}
Hawking, S.~W. 1967.
\newblock The occurrence of singularities in cosmology. {III}. {C}ausality and
  singularities.
\newblock {\em P. Roy. Soc. A-Math. Phy.\/}~{\em 300\/}(1461), 187--201.

\bibitem[\protect\citeauthoryear{Hawking}{Hawking}{1975}]{Haw75}
Hawking, S.~W. 1975.
\newblock {Particle Creation by Black Holes}.
\newblock {\em Comm. Math. Phys.\/}~{\em 43\/}(3), 199--220.

\bibitem[\protect\citeauthoryear{Hawking}{Hawking}{1976}]{Haw76}
Hawking, S.~W. 1976.
\newblock {Breakdown of Predictability in Gravitational Collapse}.
\newblock {\em Phys. Rev. D\/}~{\em 14\/}(10), 2460.

\bibitem[\protect\citeauthoryear{Hawking and Penrose}{Hawking and
  Penrose}{1970}]{HP70}
Hawking, S.~W. and R.~W. Penrose 1970.
\newblock {The Singularities of Gravitational Collapse and Cosmology}.
\newblock {\em Proc. Roy. Soc. London Ser. A\/}~{\em 314\/}(1519), 529--548.

\bibitem[\protect\citeauthoryear{Kupeli}{Kupeli}{1987a}]{Kup87b}
Kupeli, D. 1987a.
\newblock Degenerate manifolds.
\newblock {\em Geom. Dedicata\/}~{\em 23\/}(3), 259--290.

\bibitem[\protect\citeauthoryear{Kupeli}{Kupeli}{1987b}]{Kup87c}
Kupeli, D. 1987b.
\newblock Degenerate submanifolds in semi-{R}iemannian geometry.
\newblock {\em Geom. Dedicata\/}~{\em 24\/}(3), 337--361.

\bibitem[\protect\citeauthoryear{Kupeli}{Kupeli}{1987c}]{Kup87a}
Kupeli, D. 1987c.
\newblock On null submanifolds in spacetimes.
\newblock {\em Geom. Dedicata\/}~{\em 23\/}(1), 33--51.

\bibitem[\protect\citeauthoryear{Kupeli}{Kupeli}{1996}]{Kup96}
Kupeli, D. 1996.
\newblock {\em Singular Semi-{R}iemannian Geometry}.
\newblock Kluwer Academic Publishers Group.

\bibitem[\protect\citeauthoryear{Malament}{Malament}{1977}]{Malament1977TopologySpaceTime}
Malament, D. 1977.
\newblock The class of continuous timelike curves determines the topology of
  spacetime.
\newblock {\em Journal of mathematical physics\/}~{\em 18\/}(7), 1399--1404.

\bibitem[\protect\citeauthoryear{Moisil}{Moisil}{1940}]{Moi40}
Moisil, G.~C. 1940.
\newblock {Sur les g\'eod\'esiques des espaces de Riemann singuliers}.
\newblock {\em Bull. Math. Soc. Roumaine Sci.\/}~{\em 42}, 33--52.

\bibitem[\protect\citeauthoryear{Norton}{Norton}{1995}]{norton1995}
Norton, J. 1995.
\newblock Did {E}instein stumble? {T}he debate over general covariance.
\newblock {\em Erkenntnis\/}~{\em 42}, 223--245.

\bibitem[\protect\citeauthoryear{Penrose}{Penrose}{1965}]{Pen65}
Penrose, R. 1965.
\newblock {Gravitational Collapse and Space-Time Singularities}.
\newblock {\em Phys. Rev. Lett.\/}~{\em 14\/}(3), 57--59.

\bibitem[\protect\citeauthoryear{Penrose}{Penrose}{1969}]{Pen69}
Penrose, R. 1969.
\newblock {Gravitational Collapse: the Role of General Relativity}.
\newblock {\em Revista del Nuovo Cimento; Numero speciale 1\/}, 252--276.

\bibitem[\protect\citeauthoryear{Penrose}{Penrose}{1979}]{Pen79}
Penrose, R. 1979.
\newblock {Singularities and time-asymmetry}.
\newblock In {\em {General relativity: an Einstein centenary survey}},
  Volume~1, pp.\  581--638.

\bibitem[\protect\citeauthoryear{Penrose}{Penrose}{1998}]{Pen98}
Penrose, R. 1998.
\newblock {The Question of Cosmic Censorship}.
\newblock In {\em {(ed. R. M. Wald) Black Holes and Relativistic Stars}}, pp.\
  233--248. University of Chicago Press, Chicago, Illinois.

\bibitem[\protect\citeauthoryear{Schutz}{Schutz}{2012}]{schutz2012thoughtsGR}
Schutz, B. 2012.
\newblock Thoughts about a conceptual framework for relativistic gravity.
\newblock In {\em Einstein and the Changing Worldviews of Physics}, Volume~12,
  pp.\  262. Birkh\"{a}user.

\bibitem[\protect\citeauthoryear{Stoica}{Stoica}{2012a}]{Sto11f}
Stoica, O.~C. 2012a.
\newblock Analytic {R}eissner-{N}ordstr{\"o}m singularity.
\newblock {\em \href{http://stacks.iop.org/1402-4896/85/i=5/a=055004}{Phys.
  Scr.}\/}~{\em 85\/}(5), 055004.

\bibitem[\protect\citeauthoryear{Stoica}{Stoica}{2012b}]{Sto12a}
Stoica, O.~C. 2012b.
\newblock Beyond the {F}riedmann-{L}ema{\^i}tre-{R}obertson-{W}alker {B}ig
  {B}ang singularity.
\newblock {\em Commun. Theor. Phys.\/}~{\em 58\/}(4), 613--616.

\bibitem[\protect\citeauthoryear{Stoica}{Stoica}{2012c}]{Sto12e}
Stoica, O.~C. 2012c.
\newblock
  \href{http://www.degruyter.com/view/j/auom.2012.20.issue-2/v10309-012-0050-3/v10309-012-0050-3.xml}{Spacetimes
  with Singularities}.
\newblock {\em An. {\c S}t. Univ. Ovidius Constan{\c t}a\/}~{\em 20\/}(2),
  213--238.
\newblock \href{http://arxiv.org/abs/1108.5099}{arXiv:gr-qc/1108.5099}.

\bibitem[\protect\citeauthoryear{Stoica}{Stoica}{2012d}]{Sto11e}
Stoica, O.~C. 2012d.
\newblock Schwarzschild singularity is semi-regularizable.
\newblock {\em \href{http://dx.doi.org/10.1140/epjp/i2012-12083-1}{Eur. Phys.
  J. Plus}\/}~{\em 127\/}(83), 1--8.

\bibitem[\protect\citeauthoryear{Stoica}{Stoica}{2013}]{Sto13a}
Stoica, O.~C. 2013.
\newblock {\em {Singular General Relativity -- Ph.D. Thesis}}.
\newblock Minkowski Institute Press.
\newblock \href{http://arxiv.org/abs/1301.2231}{arXiv:math.DG/1301.2231}.

\bibitem[\protect\citeauthoryear{Stoica}{Stoica}{2014a}]{Sto12b}
Stoica, O.~C. 2014a.
\newblock Einstein equation at singularities.
\newblock {\em Cent. Eur. J. Phys\/}~{\em 12}, 123--131.

\bibitem[\protect\citeauthoryear{Stoica}{Stoica}{2014b}]{Sto12d}
Stoica, O.~C. 2014b.
\newblock Metric dimensional reduction at singularities with implications to
  quantum gravity.
\newblock {\em Ann. of Phys.\/}~{\em 347\/}(C), 74--91.

\bibitem[\protect\citeauthoryear{Stoica}{Stoica}{2014c}]{Sto11a}
Stoica, O.~C. 2014c.
\newblock On singular semi-{R}iemannian manifolds.
\newblock {\em Int. J. Geom. Methods Mod. Phys.\/}~{\em 11\/}(5), 1450041.

\bibitem[\protect\citeauthoryear{Stoica}{Stoica}{2015a}]{Sto15b}
Stoica, O.~C. 2015a.
\newblock Causal structure and spacetime singularities.
\newblock {\em Preprint
  \href{http://arxiv.org/abs/1504.07110}{arXiv:1504.07110}\/}.

\bibitem[\protect\citeauthoryear{Stoica}{Stoica}{2015b}]{Sto11h}
Stoica, O.~C. 2015b.
\newblock The {F}riedmann-{L}ema{\^i}tre-{R}obertson-{W}alker big bang
  singularities are well behaved.
\newblock {\em Int. J. Theor. Phys.\/}, 1--10.

\bibitem[\protect\citeauthoryear{Stoica}{Stoica}{2015c}]{Sto11g}
Stoica, O.~C. 2015c.
\newblock {K}err-{N}ewman solutions with analytic singularity and no closed
  timelike curves.
\newblock {\em U.P.B. Sci Bull. Series A\/}~{\em 77}.

\bibitem[\protect\citeauthoryear{Strubecker}{Strubecker}{1941}]{Str41}
Strubecker, K. 1941.
\newblock {Differentialgeometrie des isotropen Raumes. I. Theorie der
  Raumkurven}.
\newblock {\em Sitzungsber. Akad. Wiss. Wien, Math.-Naturw. Kl., Abt.
  IIa\/}~{\em 150}, 1--53.

\bibitem[\protect\citeauthoryear{Strubecker}{Strubecker}{1942a}]{Str42a}
Strubecker, K. 1942a.
\newblock {Differentialgeometrie des isotropen Raumes. II. Die Fl{\"a}chen
  konstanter Relativkr{\"u}mmung $K= rt- s^2$}.
\newblock {\em Math. Z.\/}~{\em 47\/}(1), 743--777.

\bibitem[\protect\citeauthoryear{Strubecker}{Strubecker}{1942b}]{Str42b}
Strubecker, K. 1942b.
\newblock {Differentialgeometrie des isotropen Raumes. III.
  Fl{\"a}chentheorie}.
\newblock {\em Math. Z.\/}~{\em 48\/}(1), 369--427.

\bibitem[\protect\citeauthoryear{Strubecker}{Strubecker}{1944}]{Str45}
Strubecker, K. 1944.
\newblock {Differentialgeometrie des isotropen Raumes. IV. Theorie der
  fl{\"a}chentreuen Abbildungen der Ebene}.
\newblock {\em Math. Z.\/}~{\em 50\/}(1), 1--92.

\bibitem[\protect\citeauthoryear{Vr\u{a}nceanu}{Vr\u{a}nceanu}{1942}]{Vra42}
Vr\u{a}nceanu, G. 1942.
\newblock {Sur les invariants des espaces de Riemann singuliers}.
\newblock {\em Disqu. Math. Phys. {B}ucure{\c s}ti\/}~{\em 2}, 253--281.

\bibitem[\protect\citeauthoryear{Zeeman}{Zeeman}{1964}]{Zeeman1964CausalityLorentz}
Zeeman, E. 1964.
\newblock Causality implies the {L}orentz group.
\newblock {\em Journal of Mathematical Physics\/}~{\em 5\/}(4), 490--493.

\bibitem[\protect\citeauthoryear{Zeeman}{Zeeman}{1967}]{Zeeman1967TopologyMinkowski}
Zeeman, E. 1967.
\newblock The topology of {M}inkowski space.
\newblock {\em Topology\/}~{\em 6\/}(2), 161--170.

\end{thebibliography}

\end{document}